\pageno=1
\baselineskip=24pt
\def\ref{\par\noindent\hang
	%indent=52pt\hangafter=1
	}
\def\lapp{\mathbin{\raise2pt \hbox{$<$} \hskip-8pt \lower3pt 
	\hbox{$\sim$}}}
\def\gapp{\mathbin{\raise2pt \hbox{$>$} \hskip-8pt \lower3pt 
	\hbox{$\sim$}}}

\def\({\left(}
\def\[{\left[}
\def\l\{{\left\{}
\def\){\right)}
\def\]{\right]}
\def\r\}{\right\}}
\def\what{\widehat}
\def\doa{\downarrow} 
\def\upa{\uparrow} 
\def\raw{\rightarrow} 
\def\bA{\bar A}
\def\bt{\bar T}
\def\bp{\bar P}
\def\am{\bar A_{\rm max}}
\def\cl{{\cal L}}
\def\cf{{\cal F}}
\def\cg{{\cal G}}
\def\de{{\rm d}}
\def\MA{M_{\rm A}^2}
\def\lb{\lambda}
\def\Lb{\Lambda}
\def\al{\alpha}
\def\be{\beta}
%
%fonts per titoli
\newif \ifamrfonts
\amrfontsfalse  % use this line if you use the cmr fonts
\ifamrfonts \font\slc=amsl10 scaled \magstep1
      \else \font\slc=cmsl10 scaled \magstep1 \fi
\ifamrfonts \font\bbbfc=ambx10 scaled \magstep3
      \else \font\bbbfc=cmbx10 scaled \magstep3 \fi
\ifamrfonts \font\bbfc=ambx10 scaled \magstep2
      \else \font\bbfc=cmbx10 scaled \magstep2 \fi
\ifamrfonts \font\bfc=ambx10 scaled \magstep1
      \else \font\bfc=cmbx10 scaled \magstep1 \fi
\
%%%\vskip 5 true cm
\vskip 3 true cm
{\bbbfc
\centerline{Matter and antimatter: the two arrows of time}
%\smallskip
%\centerline{}
}
%\medskip
%{\bbfc
%\centerline{}
%}
\bigskip
\bigskip
{\slc
\centerline{Massimo Villata}
}
\bigskip
\bigskip
%%%\bigskip
{\sl
\centerline{Istituto di Fisica Generale dell'Universit\`a, Via Pietro 
Giuria 1, I-10125 Torino, Italy}
\centerline{Osservatorio Astronomico di Torino, I-10025 Pino Torinese, 
Italy}
}
\bigskip
\bigskip
{\sl
\centerline{April 1994}
}
\bigskip
\bigskip
\bigskip
\bigskip
\noindent
%{\bf Abstract.}~ In 
{\bf 
Antiparticles may be interpreted as ordinary particles travelling 
backwards in time and the two descriptions are considered equivalent, at 
least in special relativity and relativistic quantum mechanics. It is 
suggested that, vice versa, the discovery of antimatter should be the 
confirmation that our world is ``endowed'' with two opposite time-arrows 
and such a description could be more useful and convenient from the 
point of view of the understanding of the world itself, at least for a 
simple reason: whenever phenomena are observed from a ``reference 
frame'' from which the world appears more symmetric, it is easier to 
understand the physical laws which regulate it. If, in the future, it is 
possible to discover how a macroscopic system of antimatter behaves, it 
will be also possible to confirm (or not) the ``reality'' of the two 
arrows of time.
}
%6per\bigskip\noindent
%{\sl Physics of Plasmas}, in press (issue date: June or July 1994)
\vfill\eject
It is well known that, in the special theory of relativity$^1$, the class of 
reference frames (RFs) for the description of mechanical and electromagnetic 
phenomena equivalent to a given inertial RF can be obtained by means of 
the so-called Lorentz transformations (LTs):
$$x'^\mu=\Lb^\mu_{\ \nu}x^\nu\,.\eqno(1)$$
These transformations are requested to leave invariant the quadratic form
$$\eta_{\mu\nu}\de x^\mu\de x^\nu=\eta'_{\al\be}\de x'^\al\de x'^\be\eqno(2)$$
and, from the requirement of linearity (for transforming inertial 
motion into inertial motion), it follows that
$$\eta_{\mu\nu}=\eta'_{\al\be}={\rm diag}\(+1,-1,-1,-1\)\,.\eqno(3)$$

The above equations imply that
$$\det\Lb^2=1\,,\qquad\(\Lb^0_{\ 0}\)^2\geq1\,.\eqno(4)$$
Disregarding the case of the non-proper ($\det\Lb=-1$) LTs, we are left with 
the subgroup of the proper ($\det\Lb=+1$) LTs
$$\cl=\cl^\upa\cup\cl^\doa\equiv\l\{\Lb^\upa\r\}\cup\l\{\Lb^\doa\r\}\,,
\eqno(5)$$
where $\Lb^\upa$ are the orthochronous ($\Lb^0_{\ 0}\geq+1$) LTs and $\Lb^\doa$ 
are the antichronous ($\Lb^0_{\ 0}\leq-1$) ones, which do not preserve the 
sign of time. Usually, besides the non-proper LTs, also the antichronous LTs 
are regarded as devoid of physical meaning. Some authors$^{2-10}$ have 
shown, on the contrary, that such LTs can be related to the existence of 
antimatter. We want to reconsider this issue from a slightly different point of 
view, in order to clarify some concepts which may lead to a new understanding
of antimatter. At first sight, this different interpretation might seem 
insignificant in
the today's physical perspective, but it is, instead, important in 
view of future developments of physical theories.

Consider first the quadratic form (speed of light in vacuum $c=1$)
$$\de\tau^2=\de x^\mu\de x_\mu=\eta_{\mu\nu}\de x^\mu\de x^\nu=\de t^2-\de{\bf 
x}^2\eqno(6)$$
and its canonically conjugated form 
$$m_0^2=p^\mu p_\mu=\eta_{\mu\nu}p^\mu p^\nu=E^2-{\bf p}^2\,.\eqno(7)$$
As is well known, they are invariant quantities (scalars) of the theory. We 
also know that particles travelling with speed $v=c=1$ (invariant in any RF) 
have,
consequently, $\de\tau^2=m_0^2=0$; moreover they have also the peculiarity that 
they cannot be considered at rest in any RF (their speed is always equal to 
$c$). In any other case, particles (travelling with subluminal
speeds) have $\de\tau^2>0$ and $m_0^2>0$, and, in particular, $\de\tau>0$ and 
$m_0>0$. We can infer that, for symmetry, particles endowed with 
negative proper time and rest mass should exist too. In fact, we can see them 
as antiparticles, the meaning of these negative quantities will be clarified 
in the following.

Consider now the particular antichronous transformation $-{\bf 1}\equiv
{\rm diag}\(-1,-1,-1,-1\)$ named ``total 
inversion''. It changes the sign of all components of all four-vectors.
Thus it would give the description of a particle as seen by an 
observer travelling backwards in time and with the space axes inverted, or, 
{\it vice versa}, it would provide the description of a particle having the 
opposite
``time-arrow'' and seen through a mirror with respect to us. Hence we should
see such a hypothetical particle as endowed with negative energy and opposite 
momentum direction, but it has been pointed out that we {\it are not 
able} to ``see'' such a situation, because {\it we} are constrained to go 
in {\it our} time 
direction and can see only objects which go from {\it our} past to {\it our} 
future. At this point we are faced with the so-called ``switching'' or 
``reinterpretation'' procedure (or principle)$^{11-17}$. In order 
to make the subject intuitive, consider a particle which is going from a 
given point A (on our left) to another point B (on the right) and imagine 
that we see the projection of such a simple phenomenon on a semitransparent 
screen. Now imagine that we are viewing
this travel from a RF having the time and space axes inverted, i.e.\ we put 
ourselves at the back of the screen and reproject the film backwards:
we shall see a particle going from B to A, but again from left to right.
What has happened? We ``know'' that the particle is in a state of motion 
backwards in time with negative energy and opposite momentum, but we see 
a particle moving forward in time, whose sign of the velocity is unchanged, 
and which is now
travelling from B to A. This is the effect of the ``reinterpretation'': the 
scalars ($\de\tau$, $m_0$ and, as is known, all additive charges)
have changed their signs, so that we observe
$$\de t=u_0\,\de\tau>0\,,\qquad\de{\bf x}={\bf u}\,\de\tau>0\,,\qquad
E=m_0u_0>0\,,\qquad{\bf p}=m_0{\bf u}>0\,,\eqno(8)$$
having all the right hand sides changed sign two times: first owing to the 
inversion of the 
velocity four-vector $u_\mu$ by the operation $-{\bf 1}$ and then because 
of the change in the scalars. We have also, as observed,
$${\bf v}={\de{\bf x}\over\de t}={{\bf u}\over u_0}>0\,.\eqno(9)$$

What about a particle endowed with electric charge (or any other additive 
charge)? It is well known that,
if in the first ``time reference frame'' (TRF) it is, for instance, an electron 
carrying negative charge from A to B, 
in the other TRF (after reinterpretation) we shall deal with an 
{\it antiparticle} of opposite charge, i.e.\ we observe a {\it positron} going 
from B to A. Eventually this is the ``confirmation'' that this theory including 
antichronous LTs and reinterpretation is not only a mathematical speculation 
but has its clear physical meaning, corresponding
to observable phenomena. In other words, the theory of special relativity, once 
based on the whole proper group (5) of both orthochronous and antichronous LTs, 
describes a Minkowski space-time populated by both particles and antiparticles 
and, hence, the existence of the latter could have been predicted, in purely 
relativistic, classical terms, even since 1905, exactly with the properties 
they actually exhibited when later discovered, provided that recourse to the 
switching procedure had been made. Moreover, some authors$^{2-5}$ have shown 
that the 
combination of the operation $-{\bf 1}$ with the switching procedure in special 
relativity corresponds to the CPT transformation in relativistic quantum 
mechanics, where the covariance of physical laws under such an operator is 
stated by the so-called CPT theorem.

Hence, the particle which in the first TRF carried 
positive energy and momentum from A to B (defining ``left-to-right'' as the 
positive space direction), from the other TRF is seen as a particle bearing 
positive energy and momentum from B 
to A, as requested to a ``normal'' particle travelling forward in time from 
left to right. However, such a particle has, at least formally, {\it 
negative rest mass and 
proper time} and {\it opposite additive charges}:
this is the price to pay for having ``forced'' the particle to 
travel with us in time. What about this negative rest mass? How does such a 
particle behave in a gravitational field, for example? We should expect that it 
is repulsed by ``normal'' matter of positive rest mass (i.e.\ moving forward in 
time), so as a positron is attracted by negative electric charges; this 
issue could explain the scarcity of antimatter in the universe: if 
an ``antiuniverse'' made of antimatter exists, it would be gravitationally 
repulsed by our own. In this sense, rest mass looks very much like baryonic and
leptonic numbers, and also like any other additive charge: indeed it would 
behave, at least in pair annihilation and production processes, as a quantity
conserved in time. Opening a parenthesis, we like 
to think that the ``big bang'' in which space-time and mass were created was 
nothing but the production of a universe-antiuniverse pair from a 
``non-space-time state'' of zero mass [just like a photon which is not proper 
to our space-time (indeed it has no rest RF) can create a pair], or, without 
reinterpretation, a universe which, by interacting with a ``big photon'', 
reverses its 
time-arrow. Perhaps, in turn, this ``non-space-time state'' derives from a 
universe which has returned to the ``past''. Eventually our universe might
reannihilate with its counterpart (i.e.\ return to its past and to the
big bang, for starting again), or, maybe, because of the gravitational 
repulsion, they will never meet. In this case, our universe might end by a 
``thermal death'' in a maximum entropy state, from which the only way for being 
regenerated would be to reverse its time-arrow passing through a non-space-time 
state, thus giving rise to another big bang, and so on. However, this is pure 
speculation, and we close the parenthesis.

Returning to our particles:
what about the ``reality'' of the phenomenon? Is antimatter something moving
forward in time with us, as we effectively see it, or is it normal matter going 
towards our past? Someone might
say that such a question is meaningless, since the two descriptions
(particle travelling backwards in time with negative energy and given additive 
charges, and antiparticle with 
negative mass and opposite charges going forward in time with positive energy) 
are physically equivalent, {\it at least} in special relativity and quantum 
field theory, and there are no means ({\it in the ambit of these theories}) of 
distinguishing the two situations. This is 
true, it is only a question of ``interpretation''. It is also true, as some 
authors pointed out$^{2-10}$, that we are constrained to ``explore'' space-time 
in a given time direction and, hence, it {\it seems} reasonable to perform the 
reinterpretation procedure, ``forcing'' thus all the world to go with us. We 
want to make a brief note about this, with the help of an analogy.

Imagine a situation in which an observer is constrained to explore his world 
on board of a boat unprovided with oars which is thus compelled to go 
downstream on a river or a canal. He is also bound to watch forward for some 
strange reason. This man understands that he is not at rest with respect to 
the surrounding world because the landscape does change as time goes by. He can
thus perform a measure of his position and momentum with respect to the dry 
land, regarded as 
at rest with zero momentum. This situation is analogous to our own when we 
exchange the concept of space for that of proper time and momentum for rest 
mass. Indeed we know that we are not at rest in time and this situation 
``produces'' our mass, whereas the ``terra firma'' state (photons) has constant 
proper time and zero
mass. Suppose now that our observer meet another gentleman who is going in
the opposite direction (for instance on a parallel canal with opposite stream),
he too on his boat. Remember that our dull-witted man cannot, because of 
some trouble, turn his head and follow the motion of the stranger. Therefore he 
can do nothing but look at the queer customer in the 
rearview-mirror with which his 
boat is providentially equipped, reflecting thus before himself that 
inconceivable backward motion; here is everything straight: also the other 
wayfarer travels in the ``right'' direction, as requested by the stream of the
canal, even though, unfortunately, he who is observed in such a way turns out 
to be irremediably left-hand, right-hearted and with ``{\it Aquatic Club}'' 
written from right to left, i.e.\ member of another species, in a word, an 
``antigentleman''. This is the price to pay for having wanted to force that 
man to travel with him on the river, in the ``positive'' space direction and 
with positive momentum.
All this is analogous (exchanging space for time and momentum for energy) to 
the reinterpretation procedure we usually perform when dealing with 
antimatter. Other gentlemen travelling ``forward'' but along directions forming 
an angle with that of the observer are easily ``projectable'' on it 
(orthochronous LTs). Moreover, another strange direction exists: the 
``non-projectable'' perpendicular one, i.e.\ the one corresponding to the terra 
firma, where gentlemen have not momentum (mass) and are out of the (time)
flow, so that it can provide a useful passage for inverting direction, just 
like, for instance, an electron, for changing its time-arrow, has to deal 
with a massless and timeless photon (absorbed or emitted, depending on the TRF
from which it is observed) that represents the exchange of energy (momentum) 
between the electron (gentleman) which has impinged on the non-space-time 
state (dry land) before starting again in the opposite time (space) direction, 
where it has been already observed as an antielectron.

What we mean with this analogy is that it seems more useful and reasonable not 
to stop our knowledge at the appearance of phenomena, i.e.\ not to consider 
them
only from our RF or TRF. It is true that we are not able, at least for now, to 
change our time direction, but this does not mean that we cannot imagine to see 
things from another point of view, e.g.\ from the ``terra firma'', which 
could provide
a more convenient and symmetric description. We think that none prefer, when 
going by car, to describe the motorway as something where cars and anticars 
exist, being anticars visible only in the rearview-mirror: it could be a 
misleading description of reality, even though we are aware that we are 
forbidden to reverse our direction.

In other words, the Copernican revolution has been effected without any need 
to go onto the Sun, i.e.\ into the most convenient RF 
for the understanding of the physical laws which regulate planets motions. The 
description of the retrograde motions as seen from the Earth is equivalent but 
has been misleading up to few centuries ago, because it strengthened (or did 
not allow to get over) the 
prejudice of the Earth at the centre of the universe, which, in turn, prevented 
from discovering the celestial mechanics laws. Also the ancient Egyptians 
passed through a crisis when reaching the Euphrates which flows from 
north to south, since, before its discovery, they knew only the Nile 
and its tributaries, that,
on the contrary, all flow in the opposite direction, so that, for them, the 
words (and concepts) ``north'' (``south'') and ``downstream'' 
(``upstream'') had become merged. [In fact, in their writing, ``go 
downstream (north)'' was represented by a boat without sails, and ``go 
upstream (south)'' by a boat with sails.] Their confusion is recorded in 
the stele of Tuthmosis I, in the reference to ``that inverted water 
which goes downstream (north) in going upstream (south)''$^{18,19}$. In 
other words, the ancient Egyptians discovered a sort of ``antiwater''.
We mean that it could be misleading to name antimatter (which travels forward 
in time) that normal matter which is going backwards in time, even though we 
seem constrained to follow our time-arrow and, consequently, such a description 
could seem more appropriate and simpler. On the contrary, we think that a 
description wherein we need a lower number of kinds of particles and in which 
the world is more symmetric should be the most convenient from the point of 
view of the understanding of the world itself (just like to put, ideally, 
ourselves on the Sun, i.e.\ in the more symmetric situation, was so useful and 
fruitful). 

As an example of what we mean when speaking of not considering
phenomena only from our TRF and of imagining to see things from another 
point of view, consider for a moment the well known Wheeler--Feynman 
absorber theory$^{20-23}$ (which proposes that an accelerated charge will not 
radiate unless there is to be absorption at some other distant place and 
future time) and, in particular, the pertaining Lewis' paradox$^{21}$:
``The light coming from a distant star is absorbed, let us say, by a molecule 
of chlorophyl which has recently been produced in a living plant. We say that 
the light from the star was on its way toward us a thousand years ago. What 
rapport can there be between the emitting source and this newly made molecule 
of chlorophyl?''. Namely, the {\it 
future} behaviour of a distant absorber seems to determine the {\it 
past} event of radiation. The conflict between this view and common 
sense (Lewis wrote: ``Such an idea is repugnant to all our notions of 
causality and temporal sequence.'') is due only 
to the fact that we do not consider the ``viewpoint'' of the photon 
itself: it has constant proper time and any distance and time interval are 
reduced to zero for it; somehow it ``occupies'' the whole space at every 
time in a sort of single event; it does not need to know where and when 
it will be absorbed, since all happens here and now. It is just our 
misleading space-time view which is affected by the paradox. In the same 
way even the wave-particle dualism can be seen as a ``distortion'' due
to our different point of view: since the photon (or the electromagnetic 
wave) ``pervades'' all the space, we can describe it as something 
expanding spherically around the emitter when considering its 
propagation in space, but it is more convenient to deal with a particle 
when it is absorbed somewhere. Again, all this does not matter to the 
photon which not even knows it is travelling with an oddly invariant speed 
that is only due to our necessity of splitting its existence in space 
and time at a rate inherent in the space-time itself.

Moreover, the long 
standing problem of advanced and retarded radiations$^{22-30}$ acquires new 
significance. The advanced solutions to Maxwell equations could 
merely be the retarded ones as seen from the other TRF, so as the 
positrons are the backward-running electrons: what is emitted in a TRF 
is seen as being absorbed in the other one. Another possibility is that 
we ``see'' only the retarded radiations just because we are travelling 
forward in time and the advanced ones are those which would be 
seen from the other TRF. Otherwise the backward-moving matter could not 
be visible, since its emitted photons becomes absorbed ones, and a 
hypothetical ``antistar'' would appear as a sort of black hole sucking 
radiation in rather than supplying it\dots but this is another odd and 
complicated story. On the contrary, if the advanced solutions to 
Maxwell equations have this meaning, the backward-travelling observers can 
see the star emitting the usually unobserved advanced radiation which, 
for them, becomes a retarded one. In this case, the antistar would 
appear, instantaneously, as a normal star, but, as time goes by, we could 
see its backward evolution in a universe which is going towards a big 
crunch. 

The photon, in its ``single-event interaction'', does not 
distinguish between future and past, and we, as space-time observers, 
must ``spread'' such an interaction over finite space and time 
intervals, towards the future, since we are going there, and towards 
{\it our} future when the emission comes from ``antimatter'', by means 
of its advanced radiation, since we see this one as endowed with positive 
energy. 
%(Note that, when dealing with photons, we have not at our 
%disposal proper time and mass which can change sign in view of the 
%switching procedure.)
In few words, the advanced radiation would be the one absorbed 
by backward-running receivers and this could be the explanation of why 
we do not usually see it. 
%This fact could, even when considering 
%radiating antimatter, make the two description (antimatter and 
%backward-travelling matter) equivalent.

We want to conclude with a question. Is it completely true that the two 
alternative descriptions (antimatter and backward-going matter) are 
equivalent at all? It is true in the ambit of 
special relativity and quantum field theories, but\dots what about 
thermodynamics? We are not able to see (or to produce) large collections of 
antiparticles, up to now we only ``see'' elementary objects. Maybe, in the 
future, we shall be able to observe an isolated macroscopic system of 
antimatter (e.g.\ an ``antigas'' made of ``antimolecules'') and to discover 
how it behaves. If we observe, for instance, an increase of its entropy, we 
shall be allowed to forget this discussion. If the entropy, on the contrary, 
happens to 
decrease, at least, we shall not cut the figure of the ancient Egyptians.
\vfill\eject\noindent
{\bfc REFERENCES}
\medskip\noindent
\item{1.} Einstein, A.\ {\sl Ann.\ Phys.\ (Leipzig)} {\bf 17}, 891--921 (1905).
\item{2.} Recami, E.\ \& Mignani, R.\ {\sl Riv.\ Nuovo Cimento} {\bf 4}, 
209--290 (1974); Erratum, {\bf 4}, 398 (1974). 
\item{3.} Mignani, R.\ \& Recami, E.\ {\sl Nuovo Cimento A} {\bf 24}, 
438--448 (1974).
\item{4.} Mignani, R.\ \& Recami, E.\ {\sl Int.\ J.\ Theor.\ Phys.}\ {\bf 12}, 
299--320 (1975).
\item{5.} Recami, E.\ \& Ziino, G.\ {\sl Nuovo Cimento A} {\bf 33}, 
205--215 (1976).
\item{6.} Recami, E.\ {\sl Found.\ Phys.}\ {\bf 8}, 329--340 (1978).
\item{7.} Caldirola, P.\ \& Recami, E.\ in {\sl Italian Studies in the 
Philosophy of Science} (ed Dalla Chiara, M.\ L.) 249--298 (Reidel, Boston, 
1980).
\item{8.} Recami, E.\ \& Rodrigues, W.\ A.\ {\sl Found.\ Phys.}\ {\bf 12}, 
709--718 (1982); Erratum, {\bf 13}, 553 (1983).
\item{9.} Pav\v si\v c, M.\ \& Recami, E.\ {\sl Lett.\ Nuovo Cimento} {\bf 
34}, 357--362 (1982); Erratum, {\bf 35}, 352 (1982).
\item{10.} Recami, E.\ {\sl Riv.\ Nuovo Cimento} {\bf 9}, No.\ 6, 1--178 
(1986).
\item{11.} Dirac, P.\ A.\ M.\ {\sl Proc.\ R.\ Soc.\ London, Ser.\ A} 
{\bf 126}, 360 (1930).
\item{12.} St\"uckelberg, E.\ C.\ G.\ {\sl Helv.\ Phys.\ Acta} {\bf 14}, 
588--594 (1941). 
\item{13.} St\"uckelberg, E.\ C.\ G.\ {\sl Helv.\ Phys.\ Acta} {\bf 15}, 
23--37 (1942). 
\item{14.} Feynman, R.\ P.\ {\sl Phys.\ Rev.}\ {\bf 74}, 939--946 (1948). 
\item{15.} Feynman, R.\ P.\ {\sl Phys.\ Rev.}\ {\bf 76}, 749--759 (1949). 
\item{16.} Bilaniuk, O.\ M., Deshpande, V.\ K.\ \& Sudarshan, E.\ C.\ G.\ 
{\sl Am.\ J.\ Phys.}\ {\bf 30}, 718--723 (1962). 
\item{17.} Bilaniuk, O.\ M.\ \& Sudarshan, E.\ C.\ G.\ {\sl Nature} {\bf 223},
386--387 (1969). 
\item{18.} Stele of King Tuthmosis I, translated in Breasted, J.\ H.\ 
{\sl Ancient Records of Egypt} Vol.\ 2, 31 (Russel \& Russel, Inc., New 
York, 1906).
\item{19.} Csonka, P.\ L.\ {\sl Phys.\ Rev.}\ {\bf 180}, 1266--1281 
(1969).
\item{20.} Tetrode, H.\ {\sl Z.\ Phys.}\ {\bf 10}, 317--328 
(1922).
\item{21.} Lewis, G.\ N.\ {\sl Proc.\ Natl.\ Acad.\ Sci.\ U.S.A.}\
{\bf 12}, 22--29 (1926).
\item{22.} Wheeler, J.\ A.\ \& Feynman, R.\ P.\ {\sl Rev.\ Mod.\ Phys.}\ {\bf 
17}, 157--181 (1945).
\item{23.} Wheeler, J.\ A.\ \& Feynman, R.\ P.\ {\sl Rev.\ Mod.\ Phys.}\ {\bf 
21}, 425--433 (1949).
\item{24.} Stratton, J.\ A.\ {\sl Electromagnetic Theory} 428 (McGraw-Hill, New 
York, 1941).
\item{25.} Popper, K.\ {\sl Nature} {\bf 177}, 538 (1956).
\item{26.} Hoyle, F.\ \& Narlikar, J.\ V.\ {\sl Proc.\ R.\ Soc.\ London, Ser.\ 
A} {\bf 277}, 1--23 (1964).
\item{27.} Aichelburg, P.\ C.\ \& Beig, R.\ {\sl Ann.\ Phys.\ (N.Y.)} {\bf 98}, 
264--283 (1976).
\item{28.} Stephenson, L.\ M.\ {\sl Found.\ Phys.}\ {\bf 8}, 921--926 (1978).
\item{29.} Cramer, J.\ G.\ {\sl Found.\ Phys.}\ {\bf 13}, 887--902 (1983).
\item{30.} Anderson, J.\ L.\ {\sl Am.\ J.\ Phys.}\ {\bf 60}, 465--467 (1992).
\end